\documentclass[aps,pre,twocolumn,showpacs,showkeys,groupedaddress]{revtex4}
\usepackage{graphicx} 
\usepackage{rotating}
\usepackage{amssymb}      
\usepackage{amsmath}     
\usepackage{epsfig}
\usepackage[normalem]{ulem}     
\usepackage{bm}     
\usepackage{color}

\begin{document}

\title{Periodic   windows  distribution   resulting   from  homoclinic
bifurcations in the two-parameter space}

\def\active{0}

\author{R. O. Medrano-T.$^{1,2}$ and I. L. Caldas$^2$} 

\affiliation{$^1$ Departamento  de  Ci\^encias   Exatas  e  da  Terra,
Universidade  Federal de S\~ao  Paulo, Diadema, S\~ao Paulo, Brasil}

\affiliation{$^2$ Instituto de F\'isica, Universidade de S\~ao Paulo, S\~ao Paulo, Brasil}

\begin{abstract} 

Periodic  solution  parameters,  in  chaotic dynamical  systems,  form
periodic  windows with  characteristic  distribution in  two-parameter
spaces. Recently,  general properties  of this organization  have been
reported,  but a  theoretical  explanation for  that remains  unknown.
Here,  for the  first  time  we associate  the  distribution of  these
periodic  windows  with  scaling  laws based  in  fundamental  dynamic
properties.  For the  R{\"o}ssler system, we present a  new scenery of
periodic windows composed by multiple spirals, continuously connected,
converging to different points  along of a homoclinic bifurcation set.
We show that the bi-dimensional distribution of these periodic windows
unexpectedly  follows  scales given  by  the one-parameter  homoclinic
theory.  Our  result is  a strong evidence  that, close  to homoclinic
bifurcations, periodic windows are aligned in the two-parameter space.

\vglue 0.4 
truecm

\end{abstract}


\maketitle

\section{Introduction}
\begin{figure*}[t!]
  \centerline{ \includegraphics[width=\textwidth]{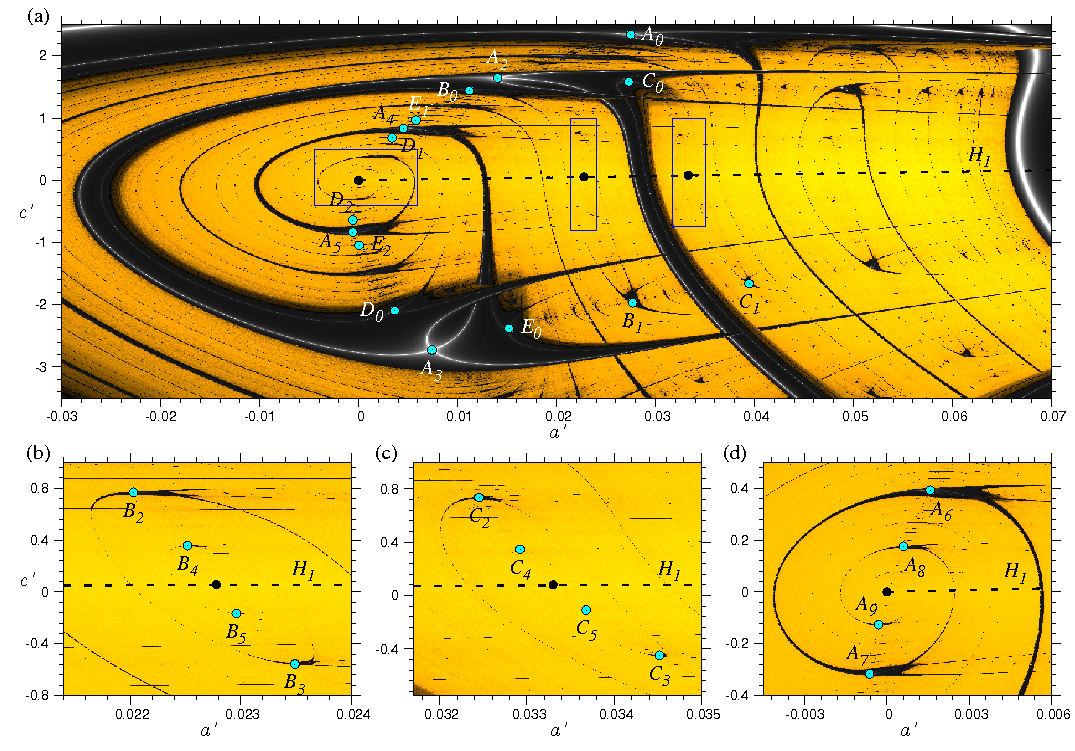} }
  \caption{(Color  online)  (a) Parameter  space  $c'\times a'$.   The
    phase space has an chaotic  attractor for parameters in yellow and
    periodic  for parameters  in gray.   The sets  $A_2$ to  $A_5$ are
    shrimps connected in a spiral way converging asymptotically to the
    onset of $H_1$ homoclinic curve (dashed line).  $B$, $C$, $D$, and
    $E$ are shrimp families emerging by flip bifurcations in $A_2$ and
    $A_3$.  The  blue square  regions are magnified  in (b),  (c), and
    (d).   The black  dots  are the  spiral  focus and  the blue  dots
    indicate the shrimp positions.}
  \label{fig-1}
\end{figure*} 

For  several smooth  nonlinear  maps and  differential
equations, stable  periodic orbits and their dependence  on the system
control parameters are well known.   The existence of these orbits can
be  properly visualized  on  a bi-dimensional  parameter space,  where
generally  we find {\it  periodic windows},  i.e., continuous  sets of
parameters, embedded into chaotic regions, for which periodical orbits
exist  \cite{Fraser1982, Gallas1993}. There  is an  intricate periodic
window,  quite general in  dynamical systems,  whose local  nature was
explained  in \cite{Fraser1982, Gallas1994}  but the  global features,
despite  the great  number of  studies  \cite{Gaspard1984, Fraser1984,
Mira1987,    Ullmann1996,   Baptista1997,    Glass2001,   Bonatto2005,
Bonatto2007,    Bonatto2008b,    Albuquerque2009,    Albuquerque2009b,
Lorenz2008, Medeiros2010}, remain not  well understood.  For local and
global  features we  mean typically  qualities  of an  isolated and  a
multiple periodic windows, respectively.   Recently, it has been found
that  these periodic windows,  baptized shrimp  \cite{Gallas1993}, are
continuously  connected along  spirals emerging  from in  a homoclinic
bifurcation point \cite{Feo2000,  Bonatto2008c} (set of parameters for
which  a bi-asymptotic  curve, the  homoclinic orbit,  converges  to a
saddle-focus  equilibrium   point).   These  spiral   structures  were
verified  experimentally in \cite{Feo2003,  Stoop2010}  and  are  also
observed in \cite{Gaspard1984, Veen2007, Albuquerque2008}.

However,  in  these researches  the  distribution  of  shrimps in  the
parameter  space   have  not  yet  been  clearly   associated  to  any
fundamental dynamical property. To accomplish this, we investigate the
relation  between  the  shrimps  and  the  homoclinic  curves  in  the
parameter  space  \cite{Medrano2008,  Medrano2010}\footnote{The  final
program     of    \cite{Medrano2008}     can    be     accessed    at:
http://www.lac.inpe.br/WSACS/imagens/ProgramaWeb.pdf}.      For    the
R{\"o}ssler  system, we  present  a new  and remarkable  two-parameter
space  scenery where  from each  shrimp emerge  infinity  spirals with
focus  in discrete  points  along of  a  homoclinic bifurcation  curve
(continuous parameter  sets for  which homoclinic orbits  exist). Each
spiral is  composed by a  shrimp family, {\it i.e.},  infinite shrimps
continuously connected in  a spiral sequence.  We show  that, even the
shrimps are a codimension-two  phenomena (two parameters are necessary
to obtain  it), they  are accumulating at  the spiral  focus following
scaling laws  predicted by  the one-parameter space  homoclinic theory
\cite{Kuznetsov2004}.  We  also show  that the reported  period adding
cascades   observed   in   shrimps  accumulations   \cite{Bonatto2007,
Bonatto2008b} is a consequence of the spiral periodic windows approach
to the homoclinic bifurcation point.

Recently was published a work \cite{Stoop2010} about scaling laws in a
electronic  homoclinic system  where  the authors  associate scales  of
tangent  bifurcation with  the shrimp  distribution.  Here  we discuss
this issue in  details and show that, in  the two-parameter space, the
scales measured in shrimps  correspond to the distance between crosses
of superstable  periodic curves. Furthermore, we call  the attention to
the  necessity  of  a  rigorous   prove  of  these  scales  in  shrimp
distributions which was done in \cite{Medrano2011}.

In  section  \ref{sec1} we  present  the  spiral  scenery of  periodic
windows,  in  \ref{sec2}  we  show  the scaling  laws  concerning  its
distribution and periodicity, and in section \ref{sec3} we present the
conclusions.

\section{\label{sec1}Shrimp distributions}

\begin{figure*}[t!]
  \centerline{ \includegraphics[width=8cm]{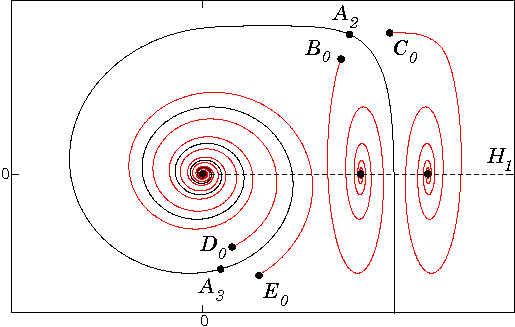} }
  \caption{(Color   online)  Schematic   scenery.   Spiral   in  black
    corresponds  to family  $A$.   In red  are  spirals emerging  from
    shrimps $A_2$ and  $A_3$. The dashed line is  the $H_1$ homoclinic
    bifurcation.   The dots along  $H_1$ are  the spiral  focus points
    while  the  points  in   the  spirals  represent  the  crosses  of
    superstables of $A_2$, $A_3$, $B_0$, $C_0$, $D_0$, and $E_0$.}
  \label{fig-2}
\end{figure*}

We consider the homoclinic R{\"o}ssler system given by
\begin{equation}
  \begin{split} 
    \dot{x}&=-y-z\\ \dot{y}&=x+ay\\ \dot{z}&=bx-cz+xz,
  \end{split}
  \label{eq-1}
\end{equation} 
where we fix  $b=0.3$ and analyze the parameter  space $c' \times a'$,
where    $a   =    a'\sin(\theta)+c'\cos(\theta)+a_0$    and   $c    =
a'\cos(\theta)-c'\sin(\theta)+c_0$  with  $a_0=0.3301$,  $c_0=4.9305$,
and   $\theta   =   88.8^{\circ}$.    For  the   region   investigated
(Fig. \ref{fig-1})  the origin phase  space ($P_0$) is  a saddle-focus
and the eigenvalues of Eqs.  (\ref{eq-1}) Jacobian matrix evaluated at
$P_0$  are $\lambda_{1,2}=\rho  \pm i\omega$  and $\lambda_3=\lambda$,
where $\lambda$, $\rho$, and $\omega$ are $\mathbb{R}^*$.

Periodic and  chaotic asymptotic  solutions of Eqs.   (\ref{eq-1}) are
determined  numerically  by evaluating  the  largest nonzero  Lyapunov
exponent $l$.  In  Fig.  \ref{fig-1}, from black to  white ($l<0$) the
periodic  orbits  increase  their  stability: in  black  these  orbits
bifurcate ($l=0$) and  in white they are superstable  ($l$ achieve the
most  negative value).  Otherwise,  from black  to yellow  ($l>0$) the
behavior is asymptotically chaotic.   The structure labeled $A_3$ is a
particular  region of  the parameter  space where  a  typical periodic
window, the  shrimp, can be  visualized.  The big shadow  region, from
which extends  four narrow  antennae, is its  {\it central  body} and
corresponds  to the  fundamental  periodic window  where  is the  {\it
  fundamental  periodic orbit} of  the shrimp  \cite{Lorenz2008}.  The
central  body  is bordered  by  tangent  bifurcations,  where we  find
chaotic regions, and by flip  bifurcations, where we find sequences of
doubled periodic regions with similar  shape to the central body.  The
white lines form  the shrimp {\it skeleton} (parameters  set where the
periodic regions are around) and represent superstable periodic orbits
with behavior  strongly attractive.  Note  that, in the  central body,
there is a cross between  two remarkable superstable curves (blue dots
in Fig. \ref{fig-1}).  We consider this point as a shrimp position and
localize it  identifying these intersections.  In  this sense, shrimps
are codimensio-two structures.

\begin{figure*}[t!]
  \centerline{ \includegraphics[width=12cm]{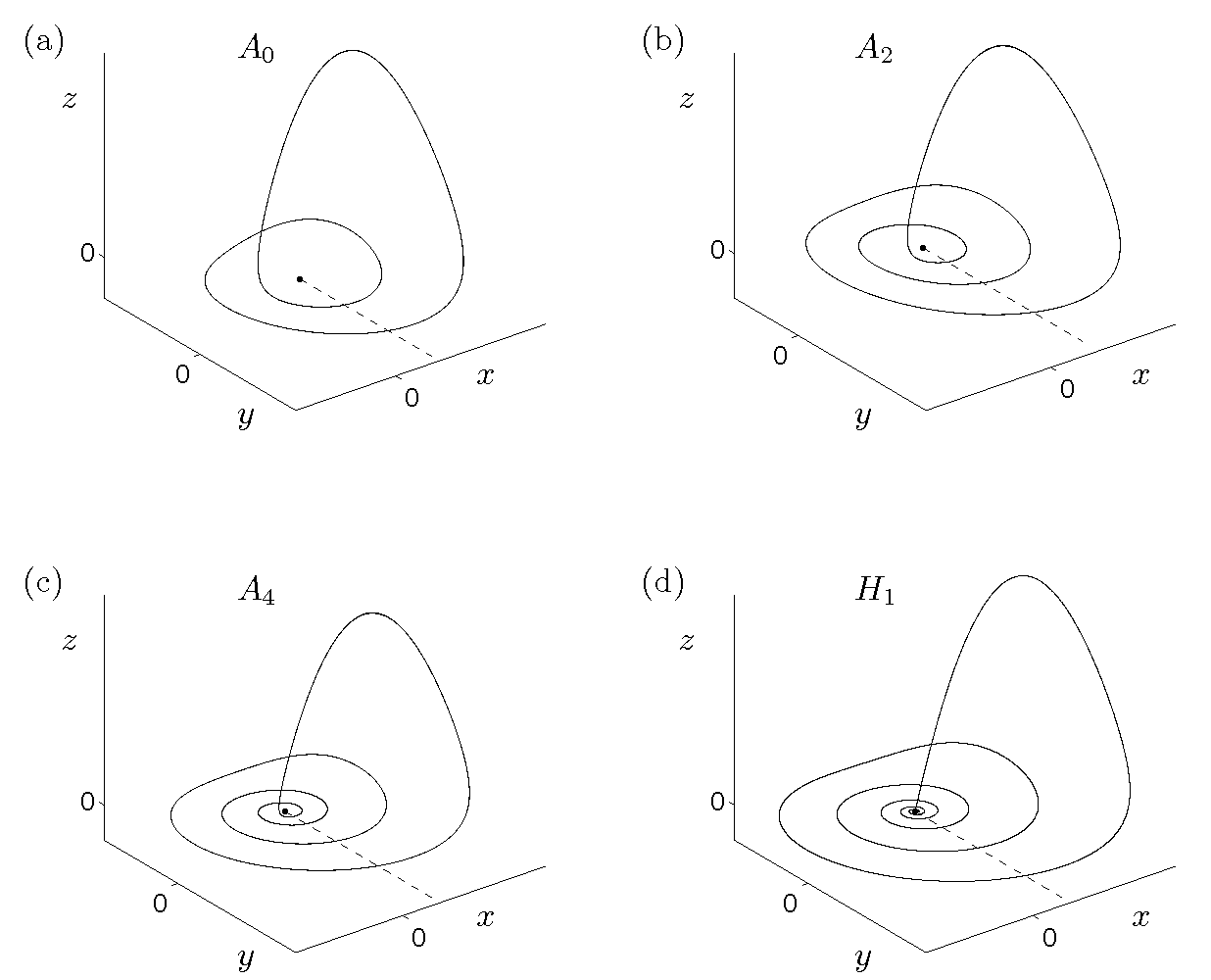} }
  \caption{Period   adding  of  $A$   shrimp  family.    The  periodic
    trajectories  cross the Poincar{\'e}  section (dashed  lines) two,
    three, four, and infinity times from (a) to (d), respectively.}
  \label{fig-3}
\end{figure*} 

To  obtain the  parameter  sets of  the  homoclinic bifurcation  $H_1$
associated with the saddle-focus point $P_0$, we determine numerically
the parameters  for which the  stable and unstable manifolds  of $P_0$
merge constituting a homoclinic  orbit (To obtain homoclinic orbits in
piecewise systems, see Ref.  \cite{Medrano2003}). The labels $A$, $B$,
$C$,  $D$, and $E$,  in Fig.   \ref{fig-1}, indicate  different shrimp
families associated  with $H_1$.  Note  that, in each  family, shrimps
are continuously connected along a spiral sequence around a homoclinic
bifurcation point [Typically, two antennae of each shrimp central body
is connected with  the previous and the next shrimp  and the other two
antennae connect  the shrimp to  the divergence region ($a'  > 0.2$)].
The index $i$,  in the label families, indicates  the element of these
sequence. The  first shrimp of the  sequence is indexed  by $i=1$, the
second by $i=2$,  and so on.  The index  $i=0$ indicates the structure
that connect  different families. Thus, $B_0$ connect  family $A$ with
family  $B$. Furthermore,  excluding the  central body  of  $A_2$, for
example, $B_0$ has the same features that characterize family $B$.

We observe that each shrimp  generates many others spiral sequences by
flip bifurcations.   The central  bodies of $A_2$  and $A_3$,  in Fig.
\ref{fig-1}(a),  bifurcate in  $B_0$, $C_0$,  $D_0$, and  $E_0$, which
connect the family $A$ with the  families $B$, $C$, $D$, and $E$.  The
family $B$  converges to  the region close  the point ($a'$  = 0.0228,
$c'$  = 0.0550)  in an  anti-clockwise  direction and  the family  $C$
converges to the region close the point ($a'$ = 0.0333, $c'$ = 0.0800)
in  a clockwise  direction. Both  focus seems  to be  along  the $H_1$
homoclinic   bifurcation   curve    [see   magnifications   in   Figs.
  \ref{fig-1}(b) and (c)].  The families  $A$, $D$ and $E$ converge to
the $H_1$ onset  point (0,0) [see Figs.  \ref{fig-1}(a)  and (d)]. The
schematic scenery is shown in  Fig. \ref{fig-2}. As far we could check,
the  same is  observed in  any element  of any  family  suggesting the
existence   of  a  fractal   structure  of   spiral  self-replications
converging  asymptotically   to  the  homoclinic   bifurcation  curve.
Moreover, we  identify shrimps concentrated  in two sets,  between the
two shrimp antennae that converge to the divergence region (as $B$ and
$C$ families),  and between  two shrimps, $i$  and $i+2$, of  the same
family (as $D$ and $E$ families).

With respect the trajectory  behavior, following continuously a shrimp
spiral sequence, we observe  that the orbit time-period grows smoothly
tending  to  infinity  close  the  homoclinic  orbit.   So  that,  the
fundamental  periodic  orbit add  one  cycle  from  $i$-shrimp to  the
$(i+2)$-shrimp forming  a period adding cascade  accumulating into the
$H_1$  curve, as  shown  in Figs.   \ref{fig-3}  (a)-(d).  The  period
adding can  be identified considering  a flat Poincar{\'e}  section in
$x=0$ with  $y<0$ close  to the plan  $xy$ (represented by  the dashed
lines).  The orbits  cross the Poincar{\'e} section two  times in (a),
three times in (b), and four times in (c).  The homoclinic orbit $H_1$
has infinite  cycles in (d). Note  that here we considered  $A_0$ as a
shrimp as discussed before.

\section{\label{sec2}Scaling laws for shrimp distributions}

Next we  analyse our numerical  results presented in the  last section
from the  homoclinic theory described by Shilnikov  theorem.  We focus
in the  characterization of  the two-parameter shrimp  structures from
these one-parameter theory.

Shilnikov  theorem  can  be  applied  to  systems  which  saddle-focus
equilibrium point have homoclinic  orbits ($\Gamma$) solutions and, in
its normal form, can be represented by
\begin{equation}
  \begin{split} 
    \dot{x'}&=\rho x'  - \omega y' + O(x',  y', z')\\ \dot{y'}&=\omega
    x' + \rho  y' + P(x', y', z')\\ \dot{z'}&=-\lambda  z' + Q(x', y',
    z'),
  \end{split}
  \label{eq-2}
\end{equation}
where   $O$,    $P$,   and    $Q$   are   analytic    functions   with
$\dot{O}(\vec{x}')=\dot{P}(\vec{x}')=\dot{Q}(\vec{x}')=0$,   for   the
equilibrium  saddle-focus point  $\vec{x}'=0$, and  $\lambda$, $\rho$,
and $\omega$ are $\mathbb{R}_+^*$.

If the saddle-focus form $\Gamma$  with saddle index $\nu<1/2$ ($\nu =
\rho/\lambda$), the  Shilnikov Theorem shows that,  close to $\Gamma$,
in  a  one-parameter  space,  there  are infinite  countable  sets  of
periodic  and  homoclinic  bifurcations  accumulating  into  $\Gamma$,
namely:  (I) Stable periodic  solutions emerge  distributed as  $S_p =
\lim_{i  \to \infty}(\mu_{i+1}  / \mu_i)  =  \exp{-(\pi \rho/\omega)}$
where $\mu_i$ and $\mu_{i+1}$  are two consecutive tangent bifurcation
parameters  \cite{Kuznetsov2004,  Glendinning1984, Gaspard1984}.   The
period difference  between two consecutive periodic  orbits is $\Delta
T=\lim_{i \to \infty}(T_{i+1} - T_i) = \pi/\omega$, where $T_i$ is the
period of the $i$-periodic  orbit; (II) Infinite classes of homoclinic
orbit,  characterized by similar  orbits, called  secondary homoclinic
orbits  [double-pulse  ($\Gamma_2$),  triple-pulse  ($\Gamma_3$),  and
  $n$-pulse  ($\Gamma_n$)] are  distributed following  $S_{\Gamma_n} =
\lim_{i  \to  \infty}(\mu'_{i+1}  /  \mu'_i) =  \exp{-(\pi  \lambda  /
  \omega)}$,  where  $\mu'_i$  and  $\mu'_{i+1}$ are  two  consecutive
homoclinic       bifurcation       parameters      of       $\Gamma_n$
\cite{Medrano2005}. The  limit $i=\infty$ is related  with the primary
single-pulse ($\Gamma_1$).

\begin{figure}[b!]
  \centerline{ \includegraphics[width=7cm]{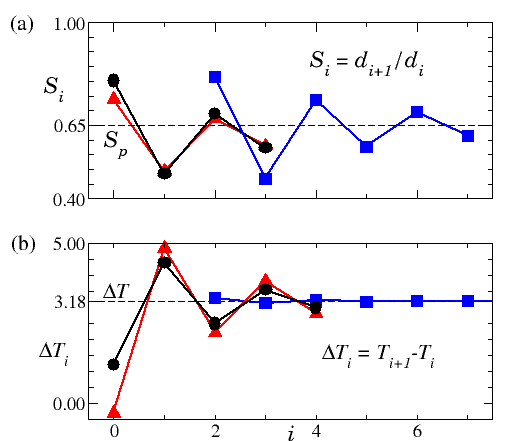} }
  \caption{(Color online)  (a) and (b) show,  respectively, measure of
    distribution and difference  time-period scalings of shrimps.  The
    dashed lines are the  theoretical values, $S_p=0.6498$ and $\Delta
    T=3.1840$,  in the  limit $i\to\infty$,  of the  family  $A$.  The
    symbols  square, triangle,  and circle  are respect  families $A$,
    $B$, and $C$.}
  \label{fig-4}
\end{figure} 

To associate  the shrimps  with this theory,  we verify  these scaling
laws in our numerical simulation.   For the family $A$, the homoclinic
orbit parameter  is $(a'_{\infty}, c'_{\infty})=(0, 0)$  for which the
eigenvalues calculated  at the fixed point $P_0$  are $\lambda_{1,2} =
0.1354 \pm  i 0.9867$  and $\lambda_3 =  -4.8713$.  Thus,  the scaling
laws are  $S_p = 0.6498$, $\Delta T=3.1840$,  and $S_\Gamma \thickapprox
2.10^{-7}$.  To measure numerically  the periodic orbit scaling $S_p$,
we  define  $S_i =  d_{i+1}/d_i$,  with  $d_i  = [(\tilde{a}_{i+1}'  -
  \tilde{a}_i')^2 + (\tilde{c}_{i+1}' - \tilde{c}_i')^2]^{1/2}$, where
$(\tilde{a}_i',\tilde{c}_i')$  is an  intersection  point between  two
superstable  curves as  defined before  the $i$-shrimp  position.  The
measured  $S_i$ for  the shrimp  family $A$  converges quickly  to the
theoretical scaling  parameter $S_p$ as shown  in Fig.  \ref{fig-4}(a)
(curve  with square  symbol).   Figure \ref{fig-4}(b)  shows that  the
measured   time-period  differences   $\Delta   T_i=T_{i+1}-T_i$  also
converge quickly to the theoretical  scaling $\Delta T$.  Thus, in the
limit  $i  \to  \infty$,  the  period  difference  between  $A_i$  and
$A_{i+2}$ is  $2\pi/\omega=6.3681$.  This interval  time corresponds to
one  periodic orbit revolution  close to  the unstable  manifold (plan
$xy$) and explains the period adding phenomenon in homoclinic systems.
It suggests  that similar mechanism  can explain the period  adding in
other  systems.   For  the  family  $B$,  we  consider  $(a'_{\infty},
c'_{\infty})=(0.0228, 0.0550)$.   The eigenvalues calculated  at $P_0$
are $\lambda_{1,2} =  0.1524 \pm i 0.9831$ and  $\lambda_3 = -4.7799$.
Thus,  the scaling  laws are  $S_p =  0.6144$, $\Delta  T=3.1956$, and
$S_\Gamma \thickapprox 2.10^{-7}$, similarly  to the family $A$.  And,
for  the family $C$,  $(a'_{\infty}, c'_{\infty})=(0.0333,  0.0800 )$,
$\lambda_{1,2} =  0.1522\pm i 0.9831$  and $\lambda_3 =  -4.7910$ with
$S_p  =0.6148   $,  $\Delta  T=3.1955$,   and  $S_\Gamma  \thickapprox
2.10^{-7}$.   The scalings  $S_i$  and  $\Delta T_i$  of  $B$ and  $C$
families  are in  agreement  to the  theoretical  estimated $S_p$  and
$\Delta  T$ as  shown in  Fig.  \ref{fig-4}(a)  and (b)  (triangle and
circle symbols for $B$ and $C$ families, respectively).  Although the
homoclinic  scaling law  parameters  is very  low  for being  verified
numerically,  we  have  observed   the  existence  of  many  different
bifurcations  $H_n$  close  to  the  primary  $H_1$  homoclinic  curve
parameter. The  scalings measured of  families $D$ and $E$  have values
very similar to the family $A$ and it was omitted in Fig. \ref{fig-4}.

We  emphasize that  it  is not  expected  that the  scaling laws  here
considered  match  with the  shrimps  scalings,  since  shrimps are  a
codimension-two phenomena.  Our  results suggest strongly that shrimps
are distributed along  lines in regions close to  the homoclinic orbit
bifurcation [See Figs. \ref{fig-1} (b)-(d)].

\section{\label{sec3}Conclusions}

We  used  the  knowledge  of  homoclinic orbits  distribution  in  the
bi-dimensional parameter space to explain the distribution of periodic
windows  in this  space. We  found infinite  periodic  structures with
spiral  shape  distributed  along  the  homoclinic  bifurcation  curve
$H_1$. Each spiral has two extremities: one is a point in $H_1$, which
is its focus, and the other is  a shrimp. In the other side, from each
shrimp derive infinite spirals with different focus along $H_1$. Thus,
the  spiral are composed  by infinite  shrimps, which  compose infinite
spirals which  have infinite shrimps which compose  infinite spirals and
son on. This characterize the fractality of this scenery of spirals.

It  is worth to  mention that,  excluding family  $A$, each  spiral is
composed by a family of shrimps that, close the $H_1$ curve, the shape
of its orbit  in the phase space approaches the  shape of an secondary
homoclinic  orbit   $\Gamma_n$  (also  called   subsidiary  homoclinic
orbit). It suggest that the real focus of each spiral is a point close
$H_1$ where  is formed  a $\Gamma_n$ orbit.  As discussed  before, the
classes  of homoclinic orbit  are organized  close $H_1$  following the
scale $S_{\Gamma_n}$.  Thus we argue  that the spiral  distribution in
the parameter space should follows the scale $S_{\Gamma_n}$.

We have  also identified  properties of the  homoclinic theory  in the
shrimps organization. We notice that this result is not expected since
the homoclinic  scalings are valid  just in a one-parameter  space and
shrimps are two-parameter structures. It indicates that shrimps, close
to  the  homoclinic  bifurcation,  are  organized along  a  line  that
intersects  $H_1$.    The  analytical  prove  of  this   is  merit  of
investigation \cite{Medrano2011}.

\vspace{.5cm} {\bf Acknowledgments}
\vspace{.3cm}

 We would like to thank Dr.  Adilson E.  Motter for important comments
 and  suggestions about  this paper,  Dr.  Manuel  A.   Mat{\'i}as for
 helpful discussions about bifurcation  theory and Prof.  Jason Gallas
 for previous discussions about  shrimp properties.  This research has
 a financial support of FAPESP and CNPq.


\end{document}